# Linear magnetization dependence and large intrinsic anomalous Hall effect in $Fe_{78}Si_9B_{13}$ metallic glasses

Weiwei Wu,[1,2] Jinfeng Li,[1,3] Zhiyu Liao,[1,2] Hongyu Jiang,[1] Laiquan Shen,[1,4] Lin Gu,[1,2] Xianggang Qiu[1,2,4] Yugui Yao,[5] Haiyang Bai,[1,2,3,4]*

[1]*Institute of Physics, Chinese Academy of Sciences, Beijing 100190, China*
[2] *School of Physical Sciences, University of Chinese Academy of Sciences, Beijing, 100049, P. R. of China*
[3]*College of Materials Science and Opto-Electronic Technology, University of Chinese Academy of Sciences, Beijing 100049, P. R. of China*
[4]*Songshan Lake Materials Laboratory, Dongguan, Guangdong 523808, China*
[5]*Key Lab of Advanced Optoelectronic Quantum Architecture and Measurement (MOE), School of Physics, Beijing Institute of Technology, Beijing 100081, China*

* Correspondence to: hybai@iphy.ac.cn

The origin of anomalous Hall effect (AHE) in ferromagnetic metallic glasses (MGs) is far from well understood due to the disordered atomic structure. Here, we report that the AHE is found in $Fe_{78}Si_9B_{13}$ MGs, and the anomalous Hall conductivity ($\sigma_{AH}$) has a linear dependence with magnetization ($M_z$), which is a feature of intrinsic mechanism contribution rather than extrinsic mechanism contribution. Moreover, the $\sigma_{AH}$ normalized by $M_z$ is independent of longitudinal conductivity ($\sigma_{xx}$), which also indicates the characteristic of non-dissipative intrinsic mechanism. This intrinsic contribution could be understood from the density of Berry curvature integrated over occupied energies proposed recently for aperiodic materials, and the linear magnetization dependence can be understood from the fluctuations of spin orientation at the finite temperature. In such case, Berry curvature is proportional to spin magnetization in different orientations. Our results indicate experimental signatures to distinguish the intrinsic mechanism in ferromagnetic MGs and confirm that the intrinsic $\sigma_{AH}$ has a large value (505-616 $Scm^{-1}$) in the $Fe_{78}Si_9B_{13}$ MGs.

# I. INTRODUCTION

The anomalous Hall effect (AHE), which exists in magnetic materials caused by broken time-reversal symmetry, has been a controversial issue since Edwin Hall's discovery in 1881 [1]. The argument is that AHE is present in almost all different types of magnetic systems, but there is currently no standard classification that covers all phenomena. In recent years, it has been believed that AHE has the same mechanism as the spin Hall effect, which can make a conversion between charge and spin current [2].

Theoretical works show three mechanisms about AHE coming from intrinsic deflection [3] and extrinsic scattering (including side jump [4] and skew scattering [5,6]) respectively. According to longitudinal conductivity $\sigma_{xx}$, the different mechanisms dominate anomalous Hall conductivity ($\sigma_{AH}$) in different regimes [7-9]: (i) in the high conductivity regime, $\sigma_{xx} > 10^6$ Scm$^{-1}$, the extrinsic skew scattering is dominant origin and scaling law acts as $\sigma_{AH} \propto \sigma_{xx}^1$; (ii) in the moderate conductivity regime, $10^4$ Scm$^{-1}$ < $\sigma_{xx}$ < $10^6$ Scm$^{-1}$, extrinsic side jump or intrinsic deflection is dominant origin and scaling law acts as $\sigma_{AH} \propto \sigma_{xx}^0$; (iii) in the lower conductivity or "bad-metal–hopping regime", $\sigma_{xx} < 10^4$ Scm$^{-1}$, some authors reported that the scaling law acts as $\sigma_{AH} \propto \sigma_{xx}^{1.6}$, which cannot be understood as any single mechanism. Therefore, in contrast to the comprehension about AHE in high and moderate conductivity regime, understanding AHE in "dirty regime" needs thorough theoretical analysis and more experimental results for many low conductivity systems.

For systems with long-range structure order, **k** is a good quantum number, and intrinsic contribution is usually understood as Berry curvature $\Omega^z(\mathbf{k})$ integrated over occupied **k** states in Brillouin zone (BZ), which can be written as [10,11]:

$$\Omega_n^z(\mathbf{k}) = -\sum_{n' \neq n} \frac{2 \operatorname{Im} \langle \psi_{n\mathbf{k}} | v_x | \psi_{n'\mathbf{k}} \rangle \langle \psi_{n'\mathbf{k}} | v_y | \psi_{n\mathbf{k}} \rangle}{(\omega_{n'} - \omega_n)^2}, \qquad (1)$$

$$\sigma_{xy} = -\frac{e^2}{\hbar} \int_{BZ} \frac{d^3 \mathbf{k}}{(2\pi)^3} \Omega^z(\mathbf{k}), \qquad (2)$$

where $\psi_{n\mathbf{k}}$ is the Bloch wave function in the nth band, $v_x$ and $v_y$ are velocity operators. While for the amorphous system without long-range translational symmetry, **k** is a bad quantum number, and the above equations are inapplicable. R. Q. Wu and F. Hellman *et al.* [12] recently suggested that it is equivalent to express the intrinsic contribution as density of Berry curvature integrated over occupied energies for aperiodic materials. The density of curvature, corresponding to the amount of Berry curvature per unit energy, is the sum of spin-orbit correlations of local orbital states and can be calculated without reference to **k** space, written as [12-14]:

$$\rho_{DOC}(\varepsilon) = \sum_{\mathbf{k}} \Omega(\mathbf{k}) \delta(\varepsilon_{\mathbf{k}} - \varepsilon), \quad (3)$$

which corresponds to the Berry curvature within an energy range bounded by $\varepsilon$ and $\varepsilon + d\varepsilon$ integrated throughout the Brillouin zone. They reported that normalized by magnetization ($M_z$) and carrier density $n^{2/3}$, $\sigma_{AH}$ of amorphous $Fe_xSi_{1-x}$ [15] and $Fe_xGe_{1-x}$ [12,16] in the high conductivity edge of the moderate conductivity regime ($10^4$ Scm$^{-1}$ < $\sigma_{xx}$ < $10^6$ Scm$^{-1}$) is independent of $\sigma_{xx}$, which is consistent with the characteristic of intrinsic mechanism. The calculated $\sigma_{AH}$ is also compatible with the experiment value [12]. These results indicate that the above theoretical model is effective to understand the intrinsic contribution of AHE in amorphous system. Moreover, Su *et al.* [17] reported that the AHE of the amorphous $Co_{40}Fe_{40}B_{20}$ thin films can be explained by the proper scaling and confirmed the existence of intrinsic contribution in the amorphous samples. Jiao *et al.* [18] reported a large inverse spin Hall effect in the Au-based and Pd-based metallic glasses (MGs), and Yang *et al.* [19] demonstrated theoretically that a topological amorphous metal phase can exist in 3D amorphous systems by calculating the Bott index, the Hall conductivity and the surface states. All above results bring opportunities to reconsider the origin of AHE in ferromagnetic MGs and explore more systems with large intrinsic contributions.

In this work, the AHE of $Fe_{78}Si_9B_{13}$, a typical ferromagnetic MG, is investigated. We find experimentally a nearly perfect linear relationship between the primitive experimental $\sigma_{AH}$ and $M_z$, which was often reported in ordered structure systems in the previous works. Moreover, the $\sigma_{AH}$ normalized by $M_z$ is independent of $\sigma_{xx}$, which is regarded as the feature of dissipationless intrinsic mechanism by Y. Yao and W-L Lee *et al.* [11,20]. With further analysis, we find this linear magnetization dependence is a feature of intrinsic contribution, and it can be obtained when extrinsic contribution is negligible or subtracted. The results can be understood qualitatively from the fluctuations of the spin orientation at the finite temperatures suggested by Zeng *et al.* [21]. In such case, Berry curvature is proportional to spin magnetization with different orientations [22]. Our results confirm that the intrinsic contribution is dominated in the AHE of $Fe_{78}Si_9B_{13}$ MGs and indicate new experimental signatures to distinguish the intrinsic mechanism for ferromagnetic MGs.

## II. EXPERIMENTAL DETAILS

The commercially available $Fe_{78}Si_9B_{13}$ MG ribbons with a uniform thickness of 25 μm were fabricated by single-roll melt-spinning method. The X-ray diffraction (XRD) with Cu $K_α$ radiation and the High-resolution transmission electron microscopy (HRTEM) were used to verify the amorphous nature of the samples. The TEM samples were carefully prepared by ion milling with 2-keV argon ions at liquid-nitrogen temperature and the HRTEM observations were conducted using a JEOL-2100 TEM. The MG ribbon was cut into a standard Hall bar by the laser beam, and the schematic is shown in inset of Fig. 1(b). The longitudinal resistivity ($ρ_{xx}$) and Hall resistivity ($ρ_{xy}$) were measured by a physical property measurement system (PPMS). The magnetic property was measured by a vibrating sample magnetometer (VSM).

## III. RESULTS AND DISCUSSION

### A. Magnetization and Resistivity

The insets in Fig. 1(a) show XRD pattern, TEM diffraction and image of the $Fe_{78}Si_9B_{13}$ MGs, and the broad diffraction peak without distinct sharp crystalline peaks indicates the amorphous nature of the MGs. Figure 1(a) shows the temperature dependence of the saturation magnetization *M* between 2 K and 350 K under a 20 kOe field, which is sufficiently high to saturate the magnetic moments in the plane of the ribbons. As the final value of saturation magnetization does not depend on the applying direction of external magnetic field, the in-plane value of *M* saturated more easily is used to represent the out-of-plane $M_z$ in this paper elsewhere. Figure 1(b) shows the *M* -$T^{3/2}$ curve, and the violet solid line is a best linear fit. The temperature range (2-350 K) is low enough compared with the Curie temperature ($T_C$ = 683K) [23] of our sample, so the best linear fit indicates spin-wave excitations, which is consistent with the results of Shen *et al*. [24]. Figure 1(c) shows in-plane *M-H* curves at the different temperatures for the MGs, and the top left inset shows detailed information at low fields. All the *M-H* curves exhibit almost perfect squareness and absence of hysteresis, indicating the excellent soft magnetic properties. The curves of $ρ_{xy}$ vs *H* perpendicular to the ribbon plane at the different temperatures are shown in Fig. 1(d). The saturation fields in Fig. 1(c) and (d) is different. Because, in Fig. 1(c), the applied field is parallel to the plane of the sample, and this direction is the easy magnetization direction, thus the saturation field is below 2kOe. While, In Fig. 1(d), the applied field is perpendicular to the plane of the sample, and this direction is the hard magnetization direction, thus the saturation field is larger than 15kOe. There are two kinds of contributions to Hall resistivity in ferromagnets, one is the ordinary Hall effect (OHE) proportional to the external magnetic field, and the other is the anomalous Hall effect (AHE) proportional to the magnetization. Then the total Hall resistivity is [25,26]:

$$\rho_{xy} = \rho_{xy}^{OHE} + \rho_{xy}^{AHE} = R_0 H + R_s M_z, \qquad (4)$$

where, $R_0$ is the ordinary Hall coefficient and $R_s$ is the anomalous Hall coefficient. From the high magnetic field data in $\rho_{xy}$-$H$ curves the carrier concentration can be determined. The anomalous Hall resistivity ($\rho_{AH}$) can be derived from the intercept by extrapolating the nearly saturated value of $\rho_{xy}$ in the $\rho_{xy}$-$H$ curves from high magnetic field to zero field. The anomalous Hall conductivity (AHC) $\sigma_{AH}$ is given by $\sigma_{AH} = -\rho_{AH}/(\rho_{AH}^2 + \rho_{xx}^2) \approx -\rho_{AH}/\rho_{xx}^2$ when $\rho_{AH}$ is much less than $\rho_{xx}$, and the anomalous Hall angle $\theta_{AH}$ is given by $\theta_{AH} = \rho_{AH}/\rho_{xx}$.

Figure 2(a) shows $R_0$ at the different temperatures, and it is positive, indicating the hole carriers are dominant. The inset shows the corresponding carrier density with the order of magnitude, $10^{22}$ cm$^{-3}$, obtained by the relation, $n_h = -1/eR_0$. Both $R_0$ and $n_h$ are slightly dependent on temperature. Figure 2(b) shows the relationship between the reduced resistivity $\rho(T)/\rho(19K)$ and temperature at different applied magnetic fields, and insets show the details in the low temperature region. From 350K to 19K, $\rho(T)/\rho(19K)$ decreases with decrease of temperature, indicating that temperature coefficient of resistivity is positive, and then a minimum resistivity is observed at 19K and $\rho(T)/\rho(19K)$ increases with further decrease of temperature. This positive temperature coefficient of resistivity has been explained in the previous works of Shen *et al*. [27]. Therefore, we mainly focus the minimum resistivity at low temperature - a common phenomenon in MGs whose origin is still an open question, and the more results and discussions on this issue can be referred to the review of Mizutani [28]. According to Shen *et al*. [27], $\rho(T)/\rho(19K)$ of Fe$_{78}$Si$_9$B$_{13}$ increases with decrease of logarithmic temperature below 19K, which is expressed as $\rho(T)/\rho(19K) = \rho_0+A\ln T$. They suggested that a spin-Kondo-type effect could be responsible for this phenomenon, which should be related with magnetic field. On the other hand, Cochrane *et al*. [29] also derived a ln$T$ term of resistivity, which is called structural Kondo effect. This effect does not refer to spin and is independent of external magnetic field. As shown in the insets of Fig. 2(b), there is no obvious difference of $\rho(T)/\rho(19K)$ with the different applied fields in our measurements. The coefficient $A$ at 0T and 7T is -(4.3±0.2) × 10$^{-4}$ (lnK)$^{-1}$, -(4.4±0.2) × 10$^{-4}$ (lnK)$^{-1}$ respectively. Therefore, it is reasonable to consider that the minimum resistivity of the sample is caused by the structure effect rather than the spin-Kondo-type effect. Whatever, more work needs to be done to explore the origin of this phenomenon.

**B. Magnetoresistance**

Figure 2(c) shows the magnetoresistance (MR) expressed as $\frac{\Delta\rho(H,T)}{\rho(T)} = \frac{\rho(H,T)-\rho(0,T)}{\rho(0,T)}$ of the Fe$_{78}$Si$_9$B$_{13}$ at the different temperatures, and the inset shows the MR for the geometry "$H(x)\|I(x)$". The MR for the geometry "$H(x)\|I(x)$" has been studied in the previous works of A. K. Majumdar *et al*. [30,31] and Yang *et al*. [32]. The main conclusions are that the MR at low field is positive and correlated with wall displacements and domain rotation processes [32], while MR at high field is negative due to less electron-magnon scattering [30,31]. In addition, S. N. Kaul *et al*. [33] has explained the composition dependence of the spontaneous resistivity anisotropy (SRA) in terms of the two-current conduction model. Therefore, the two-current conduction model may also be responsible for the MR at low field, but it needs further theoretical and experimental proof. Our measurements repeat the above results, but we also find some negative components of MR at 10-100K below saturation, which may be correlated with the negative MR for the geometry "$H(y)\perp I(x)$" [30-32] or some unexpected byproducts. For MR at out of the ribbon plane (($H(z)\perp I(x)$), based on the delocalization effect, Kawabata [34] derived the equations of positive magnetic conductivity. At high fields above technical saturation,

$$\Delta\sigma(H,T) = \sigma(H,T) - \sigma(0,T) = 0.918\sqrt{H}. \qquad (5)$$

In Fig. 2(d) and (e), MR at 350K and 10K at high fields is fitted with $\sqrt{H}$, and the corresponding coefficient is -(1.51 ± 0.04) × 10$^{-4}$ kOe$^{-1/2}$, -(1.63 ± 0.04) × 10$^{-4}$ kOe$^{-1/2}$ respectively. Using $\sigma(0,T) = 1/\rho(0,T)$, the coefficient at 350K and 10K in terms of conductivity is 1.21 ± 0.03 Ω$^{-1}$cm$^{-1}$kOe$^{-1/2}$, 1.36 ± 0.03 Ω$^{-1}$cm$^{-1}$kOe$^{-1/2}$ respectively, which do not meet the 0.918 Ω$^{-1}$cm$^{-1}$kOe$^{-1/2}$ of Kawabata at high fields. Moreover, due to less electron-magnon scattering at high fields [35], there is an isotropic magnetoresistance proportional to -$H$ for ferromagnets. Thus, in Fig. 2(f) and (g), MR at 350K and 10K is fitted with $H$, and the linear fit is better than that of using $\sqrt{H}$, manifesting as the Pearson's $r$ for $H$ (0.99154 at 350K and 0.99257 at 10K) is larger than that for $\sqrt{H}$ (0.98614 at 350K and 0.98674 at 10K). Therefore, MR at high fields is more likely due to less electron-magnon scattering rather than the suppressed localization effect of Kawabata [34].

**C. Anomalous Hall effect**

Usually, the exponent $n$ of the scaling law $R_s \propto \rho_{xx}^n$ is evaluated, where $R_s$ ($\rho_{AH}/M_z$) is anomalous Hall coefficient, to examine which mechanism governs the AHE. However, this method is not applicable for ferromagnetic MGs because both $R_s$ and $\rho_{xx}$ are slightly temperature dependent. On the other hand, N. Manyala et al. [36] reported that $\sigma_{AH}$ depends linearly on $M_z$ in the itinerant silicon-based magnetic semiconductor and considered it as the feature of intrinsic contribution. Subsequently, the same linear relation was reported for the $Co_2CrAl$ by L. J. Singh et al. [37], $La_{0.7}Sr_{0.3}CoO_3$ by Y. Tokura et al. [38] and $Fe_{0.8}Co_{0.2}Si$ by W. J. Jiang et al. [39], and all of these authors suggested the linear relationship is the characteristic of intrinsic mechanism. N. P. Ong et al. [40] further found a linear relationship between $\sigma_{AH}$ and $M_z$ in MnSi at temperatures $T < T_C$, and it can be expressed as:

$$\sigma_{AH} = S_H M_z, \qquad (6)$$

$$\rho_{AH} = S_H \rho_{xx}^2 M_z. \qquad (7)$$

The scale factor $S_H$ ($\sigma_{AH}/M_z$) shows independence of both temperature $T$ and magnetic field $H$. Therefore, we use the relation between $\sigma_{AH}$ and $M_z$ to distinguish the major contribution to AHE. In Fig. 3(a), we plot the total -$\sigma_{AH}$ versus $M_z$ of $Fe_{78}Si_9B_{13}$ MGs. The solid line is the best linear fit, and the near-perfect linear relationship between -$\sigma_{AH}$ and $M_z$ completely repeats the original results of N. Manyala et al. [36]. In Fig. 3(b), -$\sigma_{AH}$ normalized by $M_z$ almost remains constant with the change of $\sigma_{xx}$, which is consistent with the dissipationless intrinsic mechanism proposed by Y. Yao and W-L Lee et al. [11,20]. In addition, Fig. 3(c) shows the $S_H$ value of $Fe_{78}Si_9B_{13}$ in this work, which remains almost constant at different temperatures. $S_H$ values of $Co_3Sn_2S_2$ [41] and $Fe_{2.88}GeTe_2$ [42] are also shown as references. Compared to the $Fe_{78}Si_9B_{13}$, the $S_H$ of the $Fe_{2.88}GeTe_2$ [42] shows a poor constant. Because of the inelastic scattering [43] in $Fe_{2.88}GeTe_2$ [42], $S_H$ reduces moderately with the increase of temperature. Therefore, this result implies that the effect of inelastic scattering on our sample is negligible. It can be seen from Fig. 3(d) that the experimental value $\rho_{AH}$ can be well explained by the simple scaling Eq. (7), indicating that $R_s$ is equal to $S_H\rho_{xx}^2$ and also satisfies the scaling law of the intrinsic mechanism. For the $Co_3Sn_2S_2$, the first experimentally confirmed ferromagnetic Weyl semimetal [44], Wang *et al*. [41] claimed that the dependence between $\sigma_{AH}$ and $M_z$ is linear, providing the skew scattering term is subtracted. The linear relationship between $\sigma_{AH}$ and $M_z$ is indeed an experimental feature of the intrinsic mechanism. Therefore, our results suggest that the major contribution of AHE in $Fe_{78}Si_9B_{13}$ MGs is intrinsic.

Previous reports suggested that only extrinsic contribution can be ignored or subtracted then the linear magnetization dependence can be obtained. Our experimental results show that there is a nearly perfect linear relationship between $\sigma_{AH}$ and $M_z$, which confirms the inherent characteristics of AHE. However, as far as we know, there is currently no theory to explain this phenomenon. Therefore, the qualitative discussions will be made below, which may be helpful to a full explanation of this phenomenon in MGs in the future. In general, a linear dependence on the magnetization is obtained from Karplus-Luttinger theory when spin-orbit coupling (SOC) is treated as a linear perturbation [45], but this is not suitable for BCC Fe, because the SOC in iron cannot be treated accurately in a perturbative manner, according to Y. Yao *et al*. [11]. Thus, it is reasonable to believe that the above situation exists in $Fe_{78}Si_9B_{13}$, a high concentration iron glass alloy. On the other hand, Zeng *et al*. [21] suggested this linear relationship can be understood from the fluctuations of the spin orientation at finite temperatures, which means that the spin quantization axis is rotated away from the $z$ axis to a direction defined by polar angles ($\theta$, $\varphi$), in such case, Berry curvature is proportional to the spin magnetization at every angel [22], and then the intrinsic $\sigma_{AH}$ along $z$ axis is averaged over the azimuth angle $\varphi$ and proportional to $M_z$ at different temperatures. In the above case, the enhanced long spin wave fluctuations are expressed as $M$-$T^2$, according to L. Taillefer *et al*. [46], while in our case the independent spin wave excitation is expressed as $M$-$T^{3/2}$. However, both of them originate from the thermal fluctuations of the spins, which have two similar features [21]: (i) In a certain temperature range, the magnitude of spin polarization remains unchanged, but its direction deviates from the average direction due to thermal fluctuation; (ii) The typical wavelength of spin wave is much larger than the mean free path, so that the local approximation is valid. The only difference is that the intensities of fluctuations for $M$-$T^2$ are stronger. Thus, the theory proposed by Zeng *et al.* [21] with some extensions can explain the present results well. In general, we suggest that the linear magnetization dependence of primitive experiment $\sigma_{AH}$ is due to the dominance of intrinsic contribution rather than extrinsic contribution. Meanwhile, the further theoretical calculation is needed, in order to understand this linear magnetization dependence comprehensively in MG system.

The maximum experimental values of $\sigma_{AH}$ and $\theta_{AH}$ at low temperatures are 616 Scm$^{-1}$, 7.4% respectively. $\sigma_{AH}$ is larger than the candidate topological semimetal compounds GdPtBi [47] ($\sigma_{AH}$ = 110 Scm$^{-1}$) and Fe$_3$GeTe$_2$ [42] ($\sigma_{AH}$ = 540 Scm$^{-1}$), but slightly less than the series of compounds TbPtBi [48] ($\sigma_{AH}$ = 744 Scm$^{-1}$), and $\theta_{AH}$ is comparable to the ferromagnetic topological semimetals Fe$_3$GeTe$_2$ [42] ($\theta_{AH}$ = 8%). In general, a large intrinsic $\sigma_{AH}$ is associated with some special band-structure characteristics, which need further band analysis and spectrum measurement to prove. Therefore, more effort is needed to clarify the exact origin of this large intrinsic AHC in our sample.

## IV. CONCLUSION

In summary, as temperature varies, we find that the original experimental $\sigma_{AH}$ shows a linear dependence on magnetization, which is a feature of an intrinsic mechanism, indicating that the extrinsic contributions are negligible. We suggest that this linear relationship can be qualitatively understood from the fluctuations of spin orientation at finite temperature, in which case, the Berry curvature is proportional to spin magnetization of different orientations. In addition, the $\sigma_{AH}$ normalized by $M_z$ is independent of $\sigma_{xx}$, which is another feature of the dissipationless intrinsic mechanism, further indicating that intrinsic contribution predominates. Also, we find a large intrinsic $\sigma_{AH}$ (505-616 Scm$^{-1}$), which requires further band analysis and spectrum measurements to clarify the exact origin. In brief, we report experimental signatures to distinguish the intrinsic $\sigma_{AH}$ in ferromagnetic MGs and find that AHC has a large intrinsic contribution.

## ACKNOWLEDGMENTS

The authors thank H. M. Weng, S. Yang, S. Sun, S. Zhang, and H. P. Zhang for illuminating discussions. This research was supported by the Strategic Priority Research Program of the Chinese Academy of Sciences (Grant No. XDB30000000), National Key Research and Development Plan (Grant No. 2018YFA0703603, 2020YFA0308800), National Natural Science Foundation of China (Grant No. 11790291, No. 61999102, No. 61888102, No. 51871234, No. 51971238 and No. 12061131002，No. 52001220) and Natural Science Foundation of Guangdong Province (Grant No. 2019B030302010).

Reference

[1]   E. Hall, Philos. Mag. **12** (1881).
[2]   J. Sinova, S. O. Valenzuela, J. Wunderlich, C. H. Back, and T. Jungwirth, Rev. Mod. Phys. **87**, 1213 (2015). https://doi.org/10.1103/RevModPhys.87.1213
[3]   R. Karplus and J. M. Luttinger, Phys. Rev. **95**, 1154 (1954). https://doi.org/10.1103/PhysRev.95.1154
[4]   L. Berger, Phys. Rev. B **2**, 4559 (1970). https://doi.org/10.1103/PhysRevB.2.4559
[5]   J. Smit, Physica **21**, 877 (1955). https://doi.org/https://doi.org/10.1016/S0031-8914(55)92596-9
[6]   J. Smit, Physica **24**, 39 (1958). https://doi.org/https://doi.org/10.1016/S0031-8914(58)93541-9
[7]   S. Onoda, N. Sugimoto, and N. Nagaosa, Phys. Rev. Lett. **97**, 126602 (2006). https://doi.org/10.1103/PhysRevLett.97.126602
[8]   N. Nagaosa, J. Sinova, S. Onoda, A. H. MacDonald, and N. P. Ong, Rev. Mod. Phys. **82**, 1539 (2010). https://doi.org/10.1103/RevModPhys.82.1539
[9]   X.-J. Liu, X. Liu, and J. Sinova, Phys. Rev. B **84**, 165304 (2011). https://doi.org/10.1103/PhysRevB.84.165304
[10]  D. J. Thouless, M. Kohmoto, M. P. Nightingale, and M. den Nijs, Phys. Rev. Lett. **49**, 405 (1982). https://doi.org/10.1103/PhysRevLett.49.405
[11]  Y. Yao, L. Kleinman, A. H. MacDonald, J. Sinova, T. Jungwirth, D. S. Wang, E. Wang, and Q. Niu, Phys. Rev. Lett. **92**, 037204 (2004). https://doi.org/10.1103/PhysRevLett.92.037204
[12]  D. S. Bouma, Z. Chen, B. Zhang, F. Bruni, M. E. Flatté, A. Ceballos, R. Streubel, L.-W. Wang, R. Q. Wu, and F. Hellman, Phys. Rev. B **101**, 014402 (2020). https://doi.org/10.1103/PhysRevB.101.014402
[13]  G. Y. Guo, S. Murakami, T. W. Chen, and N. Nagaosa, Phys. Rev. Lett. **100**, 096401 (2008). https://doi.org/10.1103/PhysRevLett.100.096401
[14]  C. Sahin and M. E. Flatte, Phys. Rev. Lett. **114**, 107201 (2015). https://doi.org/10.1103/PhysRevLett.114.107201
[15]  J. Karel, C. Bordel, D. S. Bouma, A. de Lorimier-Farmer, H. J. Lee, and F. Hellman, EPL (Europhysics Letters) **114**, 57004 (2016). https://doi.org/10.1209/0295-5075/114/57004
[16]  J. Karel, D. S. Bouma, C. Fuchs, S. Bennett, P. Corbae, S. B. Song, B. H. Zhang, R. Q. Wu, and F. Hellman, Phys. Rev. Materials **4**, 114405 (2020). https://doi.org/10.1103/PhysRevMaterials.4.114405
[17]  G. Su, Y. Li, D. Hou, X. Jin, H. Liu, and S. Wang, Phys. Rev. B **90**, 214410 (2014). https://doi.org/10.1103/PhysRevB.90.214410

[18] W. Jiao, D. Z. Hou, C. Chen, H. Wang, Y. Z. Zhang, Y. Tian, Z. Y. Qiu, S. Okamoto, K. Watanabe, A. Hirata *et al.*, arXiv e-prints, arXiv:1808.10371 (2018).

[19] Y.-B. Yang, T. Qin, D.-L. Deng, L. M. Duan, and Y. Xu, Phys. Rev. Lett. **123**, 076401 (2019). https://doi.org/10.1103/PhysRevLett.123.076401

[20] W. L. Lee, S. Watauchi, V. L. Miller, R. J. Cava, and N. P. Ong, Science **303**, 1647 (2004). https://doi.org/10.1126/science.1094383

[21] C. Zeng, Y. Yao, Q. Niu, and H. H. Weitering, Phys. Rev. Lett. **96**, 037204 (2006). https://doi.org/10.1103/PhysRevLett.96.037204

[22] E. Roman, Y. Mokrousov, and I. Souza, Phys. Rev. Lett. **103**, 097203 (2009). https://doi.org/10.1103/PhysRevLett.103.097203

[23] R. Sahingoz, M. Erol, and M. R. J. Gibbs, J. Magn. Magn. Mater. **271**, 74 (2004). https://doi.org/10.1016/j.jmmm.2003.09.018

[24] H. G. B. S. B. Y. W. Z. X. Pan, Acta Metall Sin **20**, 333 (1984).

[25] E. M. Pugh, Phys. Rev. **36**, 1503 (1930). https://doi.org/10.1103/PhysRev.36.1503

[26] E. M. Pugh and T. W. Lippert, Phys. Rev. **42**, 709 (1932). https://doi.org/10.1103/PhysRev.42.709

[27] B.-g. Shen, H.-q. Guo, H.-y. Gong, W.-s. Zhan, and J.-g. Zhao, J. Appl. Phys. **81**, 4661 (1997). https://doi.org/10.1063/1.365517

[28] U. Mizutani, Prog. Mater Sci. **28**, 97 (1983). https://doi.org/10.1016/0079-6425(83)90001-4

[29] R. W. Cochrane, R. Harris, J. O. Ström-Olson, and M. J. Zuckermann, Phys. Rev. Lett. **35**, 676 (1975). https://doi.org/10.1103/PhysRevLett.35.676

[30] R. Roy and A. K. Majumdar, Phys. Rev. B **31**, 2033 (1985). https://doi.org/10.1103/PhysRevB.31.2033

[31] R. Singhal and A. K. Majumdar, Phys. Rev. B **44**, 2673 (1991). https://doi.org/10.1103/PhysRevB.44.2673

[32] S. U. Jen and S. M. Yang, J. Appl. Phys. **62**, 3323 (1987). https://doi.org/10.1063/1.339346

[33] S. N. Kaul and M. Rosenberg, Phys. Rev. B **27**, 5698 (1983). https://doi.org/10.1103/PhysRevB.27.5698

[34] A. Kawabata, Solid State Commun. **34**, 431 (1980). https://doi.org/https://doi.org/10.1016/0038-1098(80)90644-4

[35] S. B. Roy, A. K. Nigam, G. Chandra, and A. K. Majumdar, J. Phys. F: Met. Phys. **18**, 2625 (1988). https://doi.org/10.1088/0305-4608/18/12/013

[36] N. Manyala, Y. Sidis, J. F. DiTusa, G. Aeppli, D. P. Young, and Z. Fisk, Nat. Mater. **3**, 255 (2004). https://doi.org/10.1038/nmat1103

[37] A. Husmann and L. J. Singh, Phys. Rev. B **73**, 172417 (2006). https://doi.org/10.1103/PhysRevB.73.172417

[38] Y. Onose and Y. Tokura, Phys. Rev. B **73**, 174421 (2006). https://doi.org/10.1103/PhysRevB.73.174421

[39] W. Jiang, X. Z. Zhou, and G. Williams, Phys. Rev. B **82**, 144424 (2010). https://doi.org/10.1103/PhysRevB.82.144424

[40] S. H. Chun, Y. S. Kim, H. K. Choi, I. T. Jeong, W. O. Lee, K. S. Suh, Y S. Oh, K. H. Kim, Z. G. Khim, J. C. Woo *et al.*, Phys. Rev. Lett. **98**, 026601 (2007). https://doi.org/10.1103/PhysRevLett.98.026601

[41] Q. Wang, Y. Xu, R. Lou, Z. Liu, M. Li, Y. Huang, D. Shen, H. Weng, S. Wang and H. Lei, Nat. Commun. **9**, 3681 (2018). https://doi.org/10.1038/s41467-018-06088-2

[42] K. Kim, J. Seo, E. Lee, K. T. Ko, B. S. Kim, B. G. Jang, J. M. Ok, J. Lee, Y. J. Jo, W. Kang *et al.*, Nat. Mater. **17**, 794 (2018). https://doi.org/10.1038/s41563-018-0132-3

[43] J. G. Checkelsky, M. Lee, E. Morosan, R. J. Cava, and N. P. Ong, Phys. Rev. B **77**, 014433 (2008). https://doi.org/10.1103/PhysRevB.77.014433

[44] E. Liu, Y. Sun, N. Kumar, L. Muchler, A. Sun, L. Jiao, S. Y. Yang, D. Liu, A. Liang, Q. Xu *et al.*, Nat. Phys. **14**, 1125 (2018). https://doi.org/10.1038/s41567-018-0234-5

[45] P. Nozières and C. Lewiner, J. Phys. France **34**, 901 (1973). https://doi.org/10.1051/jphys:019730034010090100

[46] G. G. Lonzarich and L. Taillefer, J. Phys. C: Solid State Phys. **18**, 4339 (1985). https://doi.org/10.1088/0022-3719/18/22/017

[47] T. Suzuki, R. Chisnell, A. Devarakonda, Y. T. Liu, W. Feng, D. Xiao, J. W. Lynn, and J. G. Checkelsky, Nat. Phys. **12**, 1119 (2016). https://doi.org/10.1038/nphys3831

[48] R. Singha, S. Roy, A. Pariari, B. Satpati, and P. Mandal, Phys. Rev. B **99**, 035110 (2019). https://doi.org/10.1103/PhysRevB.99.035110

## Figure captions

FIG. 1. (color online). (a) Temperature-dependent saturation magnetization $M$. The insets show XRD pattern and TEM diffraction and image of $Fe_{78}Si_9B_{13}$ MGs. (b) $M$ - $T^{3/2}$. The violet solid line is a best linear fit. The inset shows a schematic Hall bar of the MG ribbon cut by laser beam. (c) In-plane $M$-$H$ curves at different temperature for the $Fe_{78}Si_9B_{13}$ MGs. The top left inset shows in-plane $M$-$H$ at low fields. (d) $\rho_{xy}$ vs $H$ curves at different temperature for the $Fe_{78}Si_9B_{13}$ MGs.

FIG. 2. (color online). (a) Temperature dependence of $R_0$ and the inset shows $n_h$ at different temperature. The black line is guide for eyes. (b) Temperature dependence of reduced resistivity $\rho(T)/\rho(19K)$ and the insets show linear fit between $\rho(T)/\rho(19K)$ and $\ln T$ below 19K at different magnetic field. (c) Magnetic field dependence of MR ($H(z) \perp I(x)$) and the inset shows the MR for the geometry "$H(x)\|I(x)$" at different temperature. (d), (e) Linear fit between MR and $H^{1/2}$ in high field. Pearson's $r$ at 350K and 10K are 0.98614, 0.98674 respectively. (f), (g) Linear fit between MR and $H$ in high field. Pearson's $r$ at 350K and 10K are 0.99154, 0.99257 respectively. Error bars in (a) and inset are the standard deviations of the linear fits when determining the $\rho_{AH}$.

FIG. 3. (color online). (a) Magnetization dependence of experimental -$\sigma_{AH}$. The black solid line is a best linear fit. (b) -$\sigma_{AH}/M_z$ (normalized AHC) vs $\sigma_{xx}$. The dash line is guide for eyes. (c) Temperature dependence of $S_H$. The solid line is guide for eyes. (d)

Temperature dependence of $\rho_{AH}$ and different fitting form $S_H\rho_{xx}^2 M_z$ (red dots), $S_H\rho_{xx}^1 M_z$ (blue dots). Error bars are the standard deviations of the linear fits when determining the $\rho_{AH}$.

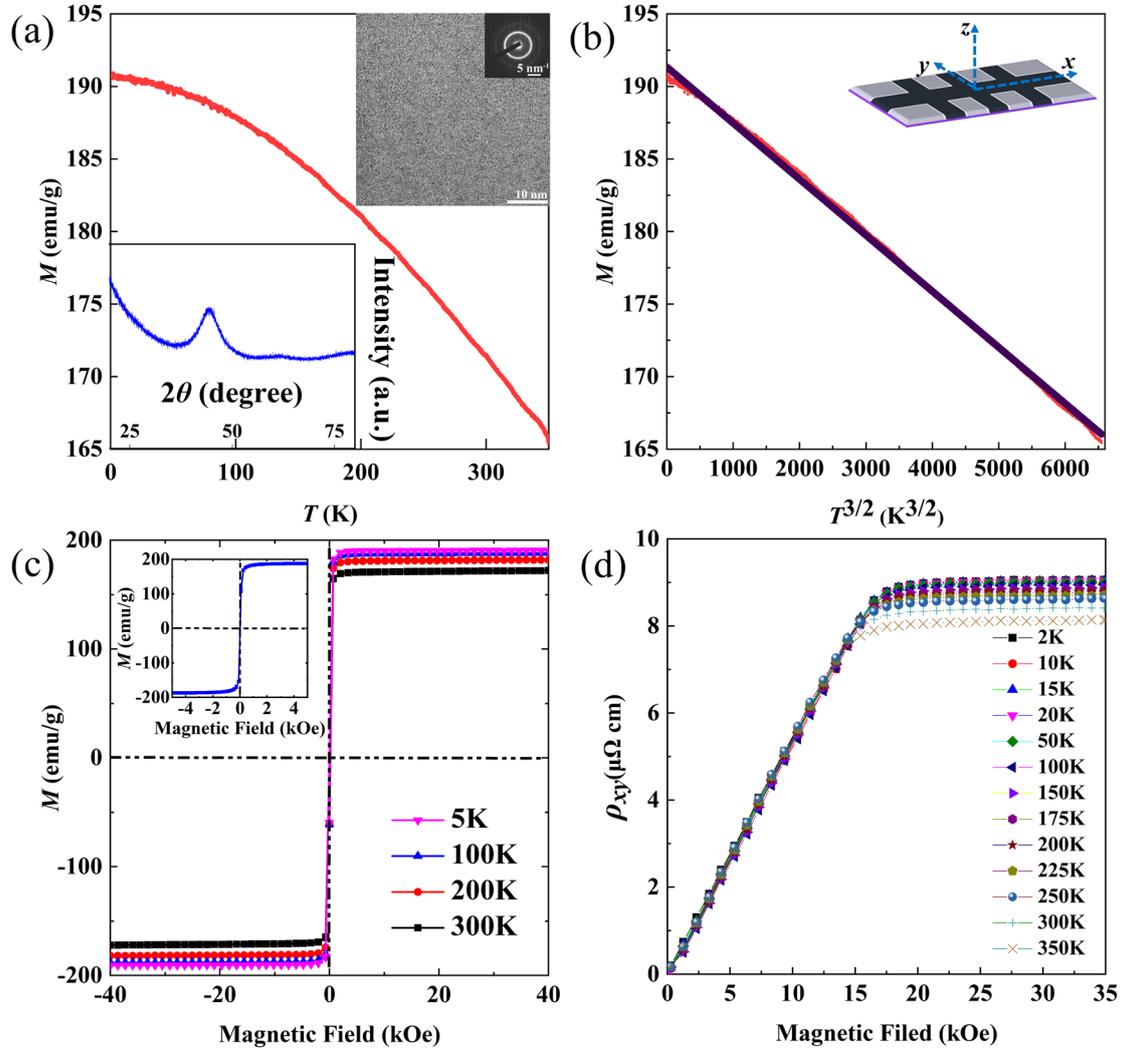

FIG. 1. Wu *et al*.

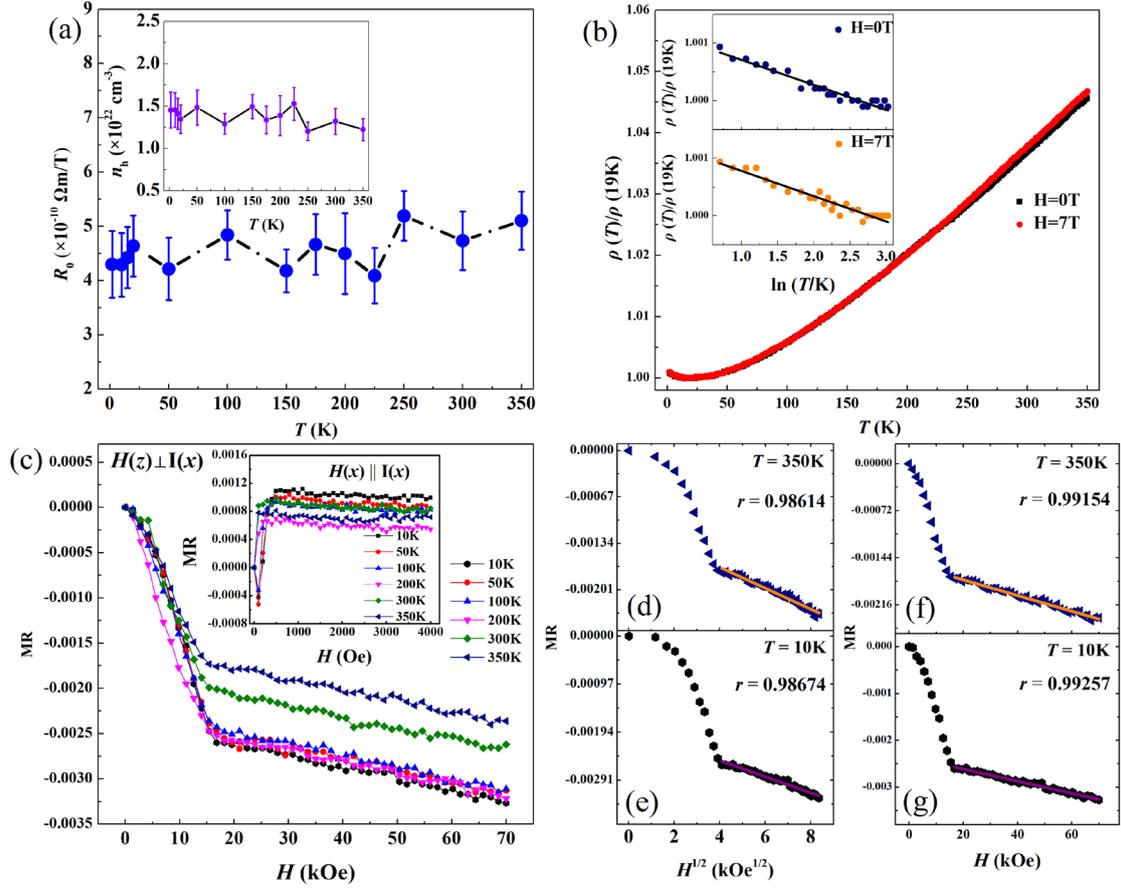

FIG. 2. Wu *et al*.

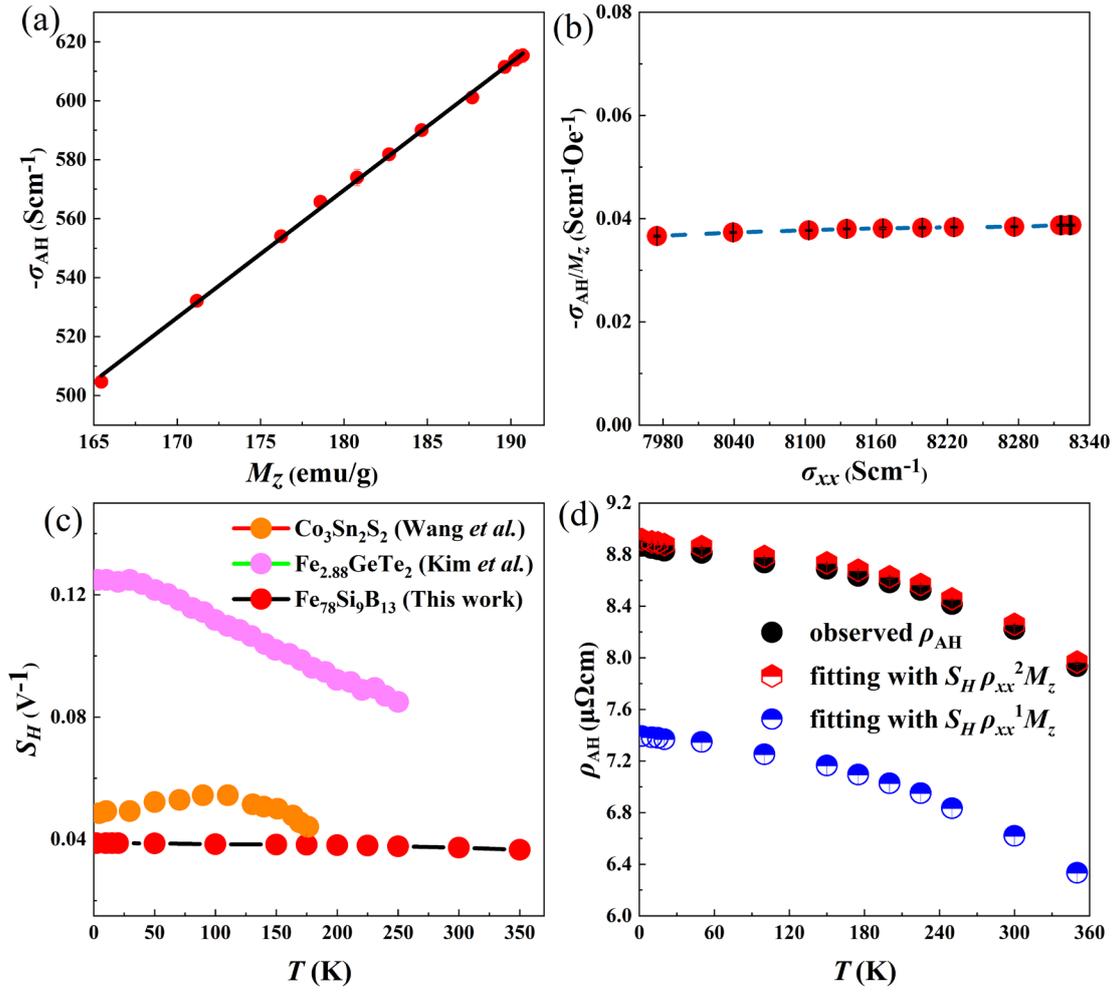

FIG. 3. Wu *et al*.